\documentclass[12pt,titlepage]{article}
\pdfoutput=1
\usepackage{setspace} 
\usepackage{amsmath}
\usepackage{amssymb}
\usepackage{graphicx,epsfig}
\usepackage[round,numbers,sort&compress]{natbib} 
\usepackage{xcolor}

\title{Differential Dynamic Microscopy: a High-Throughput Method for Characterizing the Motility of Microorganism}
\author{Vincent~A.~Martinez
	\thanks{Corresponding author.  email: vincent.martinez@ed.ac.uk},	
	Rut Besseling $^1$, 	Ottavio A. Croze $^2$, Julien Tailleur $^3$,\\
	Mathias Reufer $^1$, Jana Schwarz-Linek $^1$, Laurence G. Wilson $^{1,4}$ \\
	Martin A. Bees $^2$, Wilson C. K. Poon $^1$\\
	\\
	(1) SUPA, School of Physics and Astronomy, University of Edinburgh\\
	Mayfield Road, Edinburgh EH9 3JZ, UK
	\and
	(2) School of Mathematics and Statistics, University of Glasgow\\
	Glasgow G12 8QW, UK
	\and
	(3) Universit\'{e} Paris Diderot, MSC Sorbonne Paris Cit\'{e}\\
	UMR 7057 CNRS, F75205 Paris, France
	\and
	(4) Rowland Institute at Harward\\
         Cambridge, Massachusetts, 02142, USA}
     
\begin{document}

\maketitle

\abstract{We present a fast, high-throughput method for characterizing the motility of microorganisms in 3D based on standard imaging microscopy. Instead of tracking individual cells, we analyse the spatio-temporal fluctuations of the intensity in the sample from time-lapse images and obtain the intermediate scattering function (ISF) of the system. We demonstrate our method on two different types of microorganisms: bacteria, both smooth swimming (run only) and wild type (run and tumble)  {\it Escherichia coli}, and the bi-flagellate alga {\it Chlamydomonas reinhardtii}. We validate the methodology using computer simulations and particle tracking. From the ISF, we are able to extract (i) for {\it E.~coli}: the swimming speed distribution, the fraction of motile cells and the diffusivity, and (ii) for {\it C. reinhardtii}: the swimming speed distribution, the amplitude and frequency of the oscillatory dynamics. In both cases, the motility parameters are averaged over $\sim 10^4$ cells and obtained in a few minutes.}


\clearpage

\section*{Introduction}
The motility of micro-organisms, both prokaryotes and eukaryotes, is important in biology and medicine. For example, the virulence of the bacterial pathogen {\em Helicobacter pylori} depends on its locomotion through host epithelial mucosa \citep{Yoshiyama2000};  the phototaxis of {\em Chlamydomonas reinhardtii} and similar algae, and therefore their photosynthesis, is predicated on motility. Animal reproduction relies on motile spermatozoa. In all cases, organisms with typical linear dimension $R$ in the range $0.5\;\mu\mbox{m} \lesssim R \lesssim 10 \;\mu\mbox{m}$ swim with speeds of $v \sim 10 - 100\;\mu m/s$, so that the Reynolds number, Re~$= \rho v R /\eta$ ($\rho$ and $\eta$ are the density and viscosity of the liquid medium) are vanishingly small: in aqueous media, Re~$\lesssim 10^{-3}$. In this `creeping flow' regime, micro-organisms have evolved a variety of strategies for generating the non-reciprocating motion necessary for propulsion, e.g. by rotating or beating one or more flagella. The resulting motility phenotypes are hugely varied. Amongst these, the motility of the enteric bacterium {\em Escherichia coli} is perhaps best understood \cite{Berg2004}.

In wild type (WT) {\em E. coli}, a single cell (roughly a $1\;\mu\mbox{m} \times 2\;\mu\mbox{m}$ spherocylinder) is equipped with six to ten helical flagella (each $6-10\;\mu$m long). When these rotate counterclockwise (CCW, viewed from flagella to cell body), the individual flagella bundle together and propel the cell forward in a straight line (known as a `run'), with directional deviations brought about by orientational Brownian motion. Every second or so, one or more of the flagella reverse to clockwise (CW) rotation briefly, unbundle, and the cell undergoes large-angle re-orientation known as `tumble'. When all motors rotate CCW again, the cell begins a new run in an essentially random direction. Such `run and tumble' gives rise to a random walk, which the cell can bias by decreasing the tumble frequency when running in the direction of a favorable chemical gradient (chemotaxis).

Such detailed information in {\em E. coli} or other micro-organisms can only be obtained by single-cell tracking. On the other side,  tracking is laborious, and seldom averages over more than a few hundred cells, limiting the statistical accuracy. Moreover, since three-dimensional (3D) tracking requires specialized equipment \cite{Berg1971,Berg1972,Drescher2009}, the usual practice is to rely on measuring 2D projections in a single imaging plane, which further limits statistical accuracy because of cells moving out of the plane.

We recently proposed~\cite{Wilson2011} that differential dynamic microscopy (DDM) can be used for characterizing the motility of micro-organisms. DDM is fast, fully 3D and provides statistics from much larger samples than tracking, allowing averaging over $\sim 10^4$ cells in a few minutes, using standard microscopy imaging. We demonstrated DDM for swimming wild-type {\em E. coli}, and validated the method using simulations.

In this paper, we explain the full details of our method, discuss its limitations, and provide in-depth justification for the approximations made using computer simulation and tracking. We applied the method to a smooth-swimming (run only) mutant of  {\em E. coli}, investigated its use on the WT (run and tumble) in more detail, and extended it for the first-time to study the swimming of bi-flagellate WT  alga {\em C. reinhardtii}, a completely different microorganism in terms of time scale, length scale, and swimming dynamics. Our results should give the basis for generalizing DDM to many other biologically and medically important micro-swimmers, including spermatozoa.
\section{Differential Dynamic Microscopy} \label{sec:theory}
The key idea of DDM~\cite{Cerbino2008,Giavazzi2009,Wilson2011} is to characterize the motility of a population of particles (colloids or micro-organisms) by studying the temporal fluctuations of the local number density of particles over different length scales via image analysis. It yields the same quantity accessed in dynamic light scattering (DLS), the `intermediate scattering function' (ISF). The advantage of DDM is that the required range of length scale to study microorganism motility, such as bacteria or algae, is easily accessible in contrast to DLS~\cite{Boon1974}.

In DDM, one takes time-lapse images of particles, described by the intensity $I(\vec r, t)$ in the image plane, where $\vec r$ is pixel position and $t$ is time. As particles move, $I(\vec r, t)$ fluctuates with time. The statistics of these fluctuations contain information about the particle motions. To quantify these fluctuations, DDM measures the Differential Image Correlation Function (DICF), $g(\vec q,\tau)$, i.e. the square modulus of the Fourier transform of the difference of two images separated by $\tau$ in time
\begin{equation} \label{eq:g}
g(\vec{q},\tau)=\left\langle \left| I(\vec{q},t+\tau) - I(\vec{q},t) \right|^2 \right\rangle_{t}.
\end{equation}
Here, $\langle ... \rangle_t$ means average over the initial time $t$, and $I(\vec{q},t)$ is the Fourier transform of $I(\vec{r},t)$, which picks out the component in the image $I(\vec{r},t)$ that varies sinusoidally with wavelength $2\pi/q$ in the direction $\vec{q}$. In isotropic samples (no preferred direction of motion), the relevant variable is the magnitude $q$ of $\vec{q}$. It can be shown \cite{Cerbino2008,Giavazzi2009,Wilson2011} that $g(q,\tau)$ is related to the ISF, $f(q, \tau)$, by
\begin{equation} 
g(q,\tau) = A(q) [1-f(q,\tau)] +B(q) \label{eq:g_f},
\end{equation}
where $A(q)$ depends on the optics, the individual particle shape and the sample's structure, and $B(q)$ represents the camera noise. For independent particles, the ISF is given by \cite{Berne00}
\begin{equation}
f(q,\tau) = \left\langle e^{-i\vec{q}\cdot  \Delta \vec{r}(\tau)}\right\rangle, \label{fqt}
\end{equation}
with $\Delta \vec{r}(\tau)$ the single-particle displacement and $\langle ... \rangle$ an average over all particles.

Eq.~\ref{fqt} shows that $f(q,0) = 1$ and $f(q,\tau \rightarrow \infty) = 0$. This decay of the ISF from unity to zero reflects the fact that particle configurations (and therefore images) separated by progressively longer delay time, $\tau$, become more decorrelated due to particle motion. The precise manner in which $f(q,\tau)$ decays contains information on these motions on the length scale $2\pi/q$. The analytic form of the ISF is known for a number of ideal systems \cite{Berne00}. As an example, for identical spheres undergoing purely Brownian motion with diffusion coefficient $D$, $f(q,\tau) = e^{-Dq^2 \tau}$. On the other hand, for an isotropic population of straight swimmers in 3D with speed $v$, $f(q,\tau) = \text{sin} (qv\tau)/qv\tau$.

Figure \ref{fig:example_ISF} shows the calculated $f(q,\tau)$ for (i) diffusing spheres with about the same volume as a typical {\em E. coli} cell; (ii) isotropic swimmers with a speed distribution $P(v)$ typical of {\em E. coli}; and (iii) for a mixture of these \cite{Stock_BiophysJ78} (see later, Eq.~\ref{eq:DDM2}) at $q = 1\;\mu$m$^{-1}$. The curve for the mixture illustrates the utility of plotting the ISF against $\log \tau$: it renders obvious that there are two processes, a fast one due to swimming that decorrelates density (or, equivalently, intensity) fluctuations over $\sim 10^{-1}$~s, and a slower process due to diffusion that decorrelates over $\sim 1$~s (at this $q$). Their fractional contributions can be visually estimated to be $\approx 7:3$.
\section{Methods} \label{sec:methods}
\subsection{Samples}
\emph{E. coli} AB1157 (WT and $\Delta${\em cheY} strains \cite{mutant}) were grown in Luria-Bertani broth (LB) at 30$^\circ$C and shaken at 200~rpm, harvested in the exponential phase, washed three times by careful filtration ($0.45\;\mu$m filter) with and re-suspended in motility buffer (6.2 mM K$_2$HPO$_4$, 3.8 mM KH$_2$PO$_4$, 67 mM NaCl, 0.1 mM EDTA, pH=7.0) to optical density 0.3 (at 600 nm), corresponding to $\approx 5 \times 10^8$ cells/ml, and $\approx 0.06$\% by cell volume.  Care was taken throughout to minimize damage to flagella. A $\approx 400\;\mu$m deep flat glass cell was filled with $\approx 150\;\mu$l of cell suspension, sealed, and observed at 22$\pm 1^\circ$C.  Swimming behavior was constant over a 15 minute period.

Batch cultures of WT {\it C. reinhardtii} (CCAP 11/32B) were grown on 3N Bold's medium \cite{Harris09}, and concentrated in cotton using gravitaxis \cite{CrozeAshrafBees10}. Concentrated cell stock was diluted in growth media to optical density 0.175 (at 590 nm), corresponding to $1.4 \times 10^6$ cells/ml, and $\approx 0.002$\% by volume of cells. Cells were observed at 22$\pm 1^\circ$C in the same glass cells used for {\it E. coli} under a $600$ nm long pass filter (Cokin) to avoid stimulating the phototaxis  \cite{FosterSmyth80}. The sample dimensions are sufficiently large to avoid boundary effects and small enough to avoid bioconvection or thermal convection \cite{Drescher2009}. The algae motility was constant for 20 minutes. In all cases, we waited at least one minute before capturing images to avoid drift due to mixing flows.
\subsection{Differential dynamic microscopy}
We collected movies using a Nikon Eclipse Ti inverted microscope and a high-speed camera (Mikrotron MC 1362) connected to a PC with a frame grabber card with 1GB onboard memory. The CMOS pixel size ($14\;\mu\mbox{m}\times 14\;\mu \mbox{m}$) and magnification determine the inverse pixel size $k$ (in pixel/$\;\mu$m) in the image plane, which, together with the image size $L$ (in pixels), define the spatial sampling frequency ($q_{\rm min}=2\pi k/L$). For bacteria, $10\times$ phase-contrast movies were acquired at $L=500$, which gives $k = 0.712\;\mu \text{m}^{-1}$ and a $q$ range of $0.01 \lessapprox q \lessapprox 2.2\;\mu \text m^{-1}$. This allows the imaging of $\sim 10^4$ bacteria cells at a bulk cell density of $5 \times 10^8$ cells/ml in a $0.49\;\text{mm}^2$ field of view with a depth of field $\delta\approx 40\;\mu \text{m}$, over 38 s at a frame rate of 100 fps. For algae, $4\times$ bright-field movies were acquired at $L=500$, giving $k = 0.285\;\mu \text{m}^{-1}$ and a $q$ range of $0.004 \lessapprox q \lessapprox 0.9\;\mu \text m^{-1}$, and allowing the imaging of $\sim 10^4$ algae cells at a bulk cell density of $1.4 \times 10^6$ cells/ml in a $3.2\;\text{mm}^2$ field of view with $\delta\approx 200\;\mu \text{m}$, over 3.8 s at a frame rate of 1000 fps. We imaged at $\approx 200\;\mu \mbox{m}$ from the bottom of a $400\;\mu \mbox{m}$ thick glass capillary cell to minimize wall effects.
\subsection{Data reduction and fitting} \label{sec:DDMmethod}
\begin{figure}
   \begin{center}
      \includegraphics*[width=3.25in]{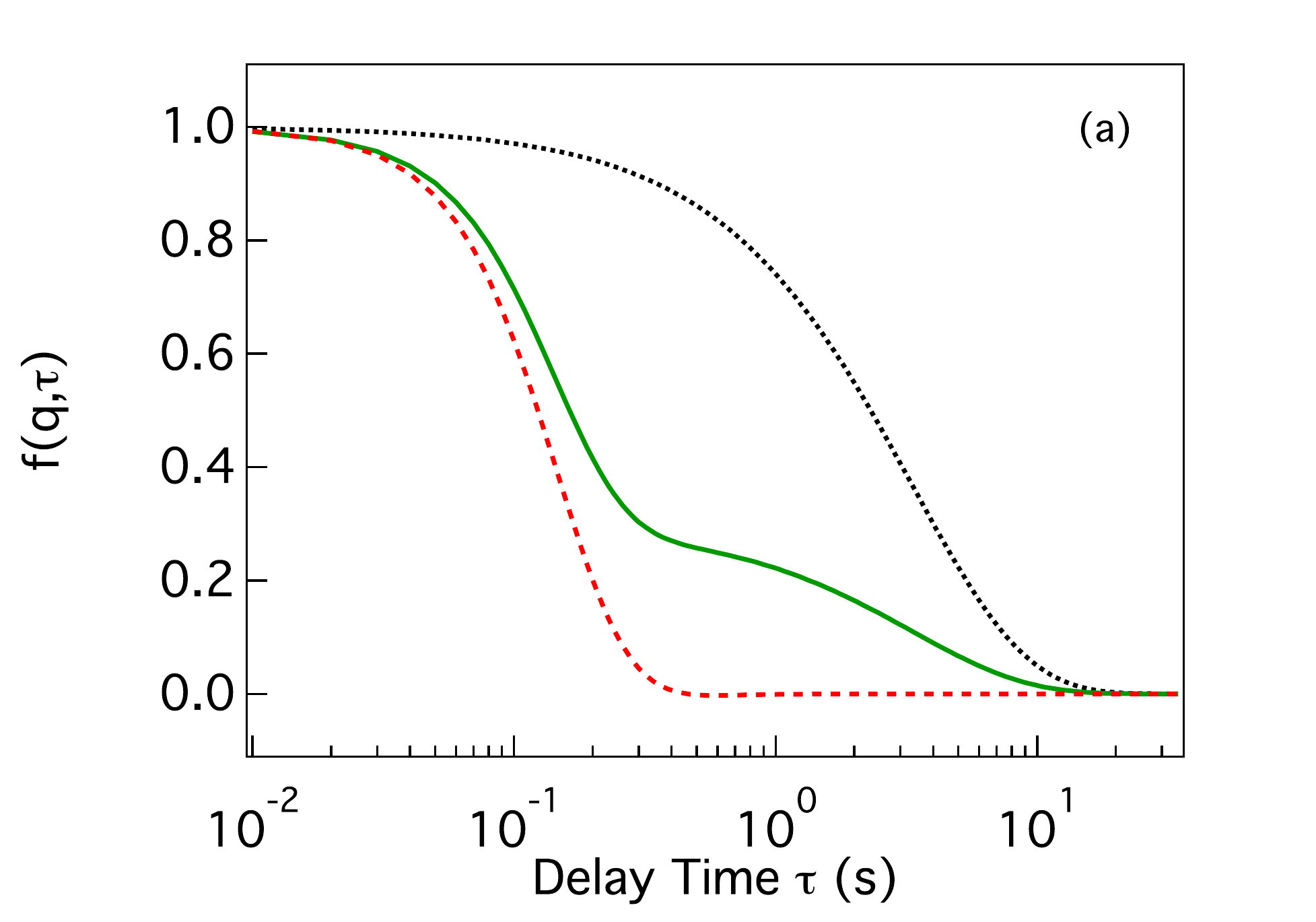}
      \includegraphics*[width=3.25in]{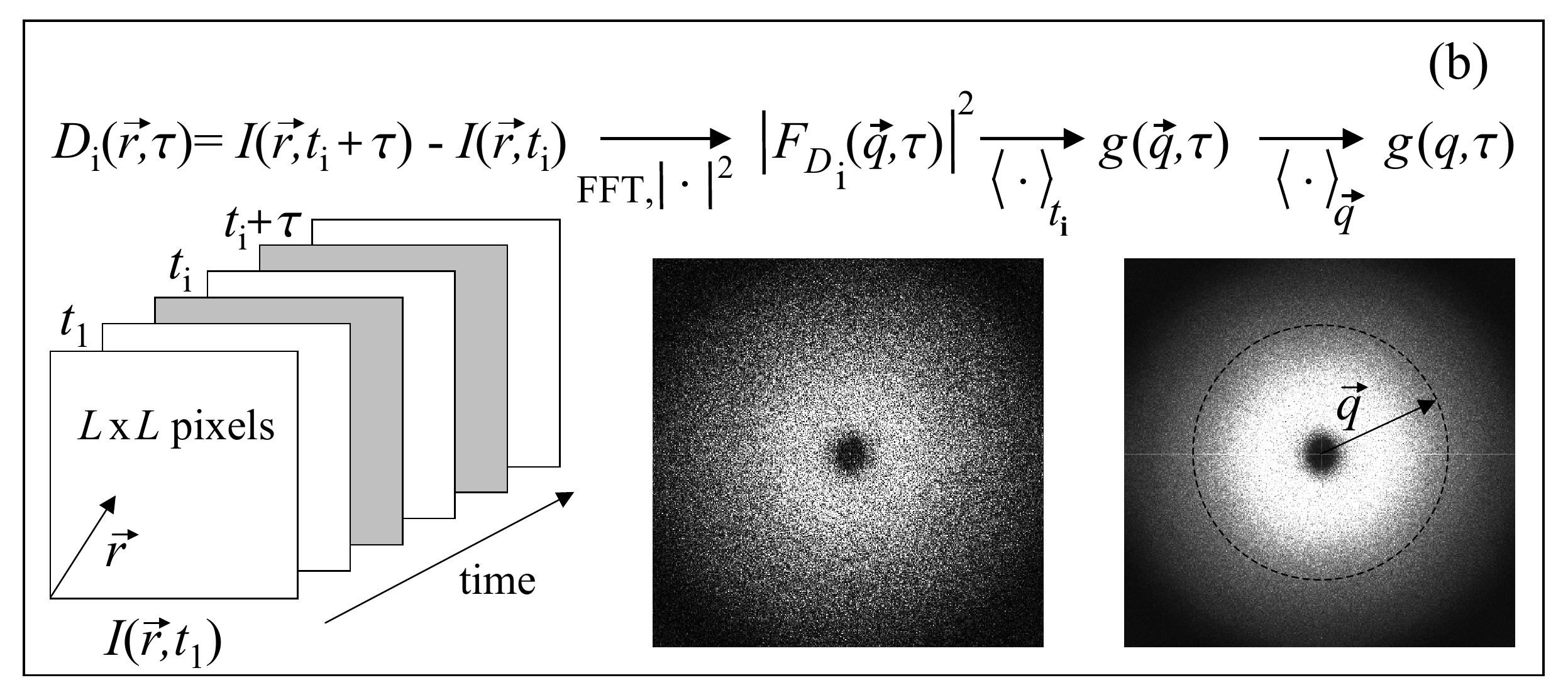}
      \caption{(a) Theoretical ISF, $f(q,\tau)$, vs $\tau$ at $q=1\;\mu {\rm m}^{-1}$, for (black dotted line) a population of diffusing spheres with $D=0.3\;\mu {\rm m}^2 {\rm /s}$, (red dashed line) a population of equivalent-size spheres swimming isotropically in 3D with a Schulz speed distribution $P(v)$ with average speed $\bar{v}=15\;\mu \mbox{m/s}$ and width $\sigma=7.5\;\mu\mbox{m/s}$, and (green line) a 30:70 mixture of diffusers and swimmers. (b) Schematic of the image processing to obtain the DICFs, $g(q,\tau)$, from the videos (left) collected in an experiment. (Middle) Non-averaged image, $|F_{D{\rm i}} (\vec{q},\tau)|^2$ and (right) averaged image, $g(\vec{q},\tau)$, over initial times $t_i$ at $\tau=0.52\;s$.}
      \label{fig:example_ISF}
   \end{center}
\end{figure}

The image processing and fitting analysis are easily automated and details are given below \cite{software}. Figure \ref{fig:example_ISF}b illustrates how we obtain the DICF, $g(q,\tau)$, from the movies. For a given delay time $\tau$, the difference images  $D_{\rm i} (\vec{r},\tau )=I(\vec{r},t_{\rm i}+\tau)-I(\vec{r},t_{\rm i}) $ are calculated for a set of $N$  different initial times $t_{\rm i}$ (typically ${\rm i} = 1, 4, 7, \ldots, 313$). After computing the fast Fourier transform, $F_{D {\rm i}} (\vec{q},\tau)$, of each $D_{\rm i} (\vec{r},\tau )$ and calculating the squared modulus, $|F_{D {\rm i}} (\vec{q},\tau) |^2$, we  average over initial times $t_{\rm i}$, giving $g(\vec{q},\tau)=\langle |F_{D{\rm i}} (\vec{q},\tau) |^2 \rangle_i$, to improve the signal-to-noise ratio (averaged image appears less grainy, Fig.~\ref{fig:example_ISF}b).

For isotropic swimmers, $g(\vec{q},\tau)$ is azimuthally symmetric and can be azimuthally averaged to give $g(q,\tau)=\langle g(\vec q,\tau) \rangle_{\vec q}$. We linearly interpolate between four adjacent points in discrete $\vec{q}$-space to find values for $g(\vec{q},\tau)$ along a circle of equidistant points with radius $q$. The finite image size causes numerical artefacts \cite{Giavazzi2009} mainly along the horizontal and vertical center lines of the image $g(\vec{q},\tau)$; these are reduced by omitting the values for $q_x=0$ and  $q_y=0$ during the azimuthal averaging. The procedure is repeated for a set of delay times $\tau$ to obtain the full time-evolution of $g(q,\tau)$. Calculations were done in LabView (National Instruments) on an 4-core PC (3 GHz Quad core, 3 GB RAM) . Processing $\lesssim 4000$ frames with $L=500$ and averaging over $\approx 100$ initial times $t_i$ take $\approx$~5 min.

We fitted independently each $g(q,\tau)$ to Eq. \ref{eq:g_f} using the appropriate parametrized model for $f(q,\tau)$. At each $q$, non-linear least-squares fitting based on $\chi^2$ minimization using the Levenberg-Marquardt algorithm and the `all-at-once-fitting' procedure in IGOR Pro (WaveMetrics) returns $A(q)$, $B(q)$ and motility parameters.
\subsection{Simulation}
We carried out Brownian dynamics simulations in 3D of non-interacting point particles (`bacteria') at a concentration and in a sample chamber geometry directly comparable to our experiments, using periodic boundary conditions to keep the bulk density of swimmers constant. Each particle has a drift velocity whose direction and magnitude were chosen from uniform and Schulz distributions, respectively. During a tumble event, a wild-type (run and tumble) swimmer undergoes standard Brownian diffusion and a new swimming direction is chosen uniformly at random after each `tumble'. The swimming speed is constant for each bacterium.

From these simulations, we constructed 2D pixelated `images'. Particles in a slice of thickness $d$, centered at $z=0$, contribute to the image. A particle at $(x,y,z)$ is `smeared' into an `image' covering the pixel containing $(x,y)$ and the 8 neighboring pixels. To define the image contrast of a bacterium, which depends on $z$, we used the experimentally measured $z$-contrast function $c(z)$ (Section \ref{section:DF}). This mimics the finite depth of field in a microscope.
\subsection{Tracking}
\label{trackmethod}
Both  experimental and simulated data were analyzed using standard particle tracking \cite{CrockerGrierJColIntSc96_tracking} giving the 2D tracks $r_{2D}(t)$. We used inverted $20\times$ videos of {\em E. coli} with bright cells of $\approx 3$ pixels on a dark background and a running average of 3 frames to improve the features' signal over noise ratio. In simulations, equivalent $10\times$ videos could be tracked due to the absence of noise. In all cases, $>400$ features were identified per frame, using only high brightness features near the focal plane. Tracking of simulated movies of non-motile (NM) or motile cells reproduced the input diffusion coefficient $D$ and swimming speed distribution $P(v)$\cite{trackfoot}. Tracking of experimental data of purely NM {\it E. coli} yields the same D as from DDM.

The analysis of (simulated or experimental) mixed populations of motile and NM cells is more challenging. We generalized a recently-proposed method \cite{Clement2011} to analyze such data. Each trajectory is split into short elementary segments of duration $\Delta t$ over which an average swimmer moves $\approx 1$~pixel. First, the mean angle $\langle |\theta| \rangle$ between successive segments is calculated; $\langle |\theta| \rangle=\pi/2$ for a random walk and $\langle |\theta| \rangle = 0$ for a straight swimmer. Then, using the trajectory's start-to-end distance $L$, duration $T$, and the mean elementary segment length $\Delta r_{2D}(\Delta t)$, we calculate the parameter $N_c=\frac{L/\Delta r_{2D}}{T/ \Delta t}$. Thus $N_c = 0$ for a random walk with $T \rightarrow \infty$ and $N_c = 1$ for a straight swimmer. Previous tracking of mixed swimmers and diffusing particles in 2D (at a wall) \cite{Clement2011} returned two well-separated clusters in the $(N_c, \langle |\theta| \rangle)$ plane, from which motile and NM populations could be separated and the respective $P(v)$ and $D$ - via fitting of the mean-squared displacement (MSD), $\langle \Delta r_{2D,NM}^2 (\tau) \rangle=4D\tau$ - could be extracted.

However, our bulk data (Section.~\ref{subsec:tracking}) show a much less well defined separation in contrast to the clear distinction in \cite{Clement2011} between motile and NM populations (specific for near wall dynamics). Therefore, we studied the dependence of motility parameters with the population selection criteria $(N_c, \langle |\theta| \rangle)$. In addition, we use another estimate for the diffusion coefficient, $D_g$, obtained by fitting the distribution of 1D displacements, $P(\Delta x_{NM}(\tau))$, to a Gaussian and using the linear increase of the variance of the fitted distribution with $\tau$ to obtain $D_g$.

Finally we tracked the {\it C. reinhardtii} videos, identifying $\approx 300$ algae per frame with $\approx 5$ pixels per cell, and applied the above track diagnostic method (using $\Delta t$ such that $\Delta r(\Delta t)\approx 1$ pixel on average) to separate `straight' tracks in the imaging plane from other tracks. Further details are given in Section~\ref{sec:algaetrack}.
\section{Smooth swimming \protect \emph{E. coli}} \label{sec:smoothswimmers}
In \cite{Wilson2011}, we demonstrated DDM using a WT strain of {\em E. coli}, in which cells `run and tumble'. We suggested that while salient features of bacterial motility could be explained ignoring the effect of tumbling, some details of the data, such as a small $q$-dependence in the fitted swimming velocity, could only be understood by taking tumbling into account. Here, we present measurements for a {\em smooth swimming} mutant of {\em E. coli}. The simplicity of the motion compared to the WT makes this mutant the ideal organism for presenting the details of DDM. We return to the WT in Section~\ref{sec:wildtype}.
\subsection{Model of $f(q,\tau)$}
In a smooth swimming (SS) mutant, each cell possesses a flagellar bundle that rotates exclusively CCW (at $\sim 100$~Hz); this propels the cell  in a straight line, but angular deviations accumulate from orientational Brownian motion. Since the whole `cell+flagella' complex must be torque free, the cell body rotates CW (at $\sim 10$~Hz). Moreover, the flagella bundle in general propels the cell off-centered, therefore the cell body `wobbles'.

To extract motility parameters from the ISF, $f(q,\tau)$, it is important to measure this function in the appropriate $q$ range. An upper bound for $q$ exists because at $q \gtrsim 2\pi/R \sim 6\;\mu{\rm m}^{-1}$, where $R \sim 1\;\mu$m is a typical cell size, both swimming and body wobble contribute to the decay of $f(q,\tau)$, so that it is impractical to extract swimming parameters cleanly in this regime. We thus need to access lower $q$, or, equivalently, larger length scales. A lower bound for $q$ is set by deviations from straight-line swimming due to Brownian orientational fluctuations and/or tumbling. For {\em E. coli}, cells run for $\sim 20\;\mu$m between tumbles; this is also the `persistence length' of the trajectory of SS cells due to orientational fluctuations. Thus, we do not want to probe much below $q \sim 0.5\;\mu{\rm m}^{-1}$.

Within the range $0.5\;\mu{\rm m}^{-1} \lesssim q \lesssim 6\;\mu {\rm m}^{-1}$, it is possible to model a population of swimming {\em E. coli} as straight swimming particles with a speed distribution $P(v)$ and uniformly distributed directions. Each particle also undergoes Brownian motion, with diffusivity $D$. To model a natural population, which inevitably contain non-motile cells, we specify that only a fraction $\alpha$ of the particles are swimming. The resulting ISF has been derived before \cite{Stock_BiophysJ78}:
\begin{equation}\label{eq:DDM2}
f(q,\tau)\!=(1-\alpha)e^{-q^2D\tau} + \alpha e^{-q^2D\tau}\!\! \int_0^\infty \!\!\!\!\!P(v)\frac{\text{sin}(qvt)}{qvt}dv.
\end{equation}
To use Eq.~\ref{eq:DDM2} to fit experimental data, we need a parametrized form for $P(v)$. Limited previous data \cite{Nossal1971,Stock_BiophysJ78} suggest that $P(v)$ is peaked. Using a Schulz (or generalised exponential) distribution 
\begin{equation}\label{eq:schultz}
P(v) = \frac{v^Z}{Z!} \left(\frac{Z+1}{\bar{v}}\right)^{Z+1} \text{exp}\left[-\frac{v}{\bar v}\left(Z+1\right)\right],
\end{equation}
where $Z$ is related to the variance $\sigma ^2$ of $P(v)$ via $\sigma =\bar{v}(Z+1)^{-1/2}$, leads to the following analytical solution of the integration in Eq.~\ref{eq:DDM2} \cite{Pusey_JChemPhys1984}
\begin{equation}\label{eq:analytical_integral}
\int_0^\infty \!\!\!P(v) \frac{\text{sin}(q v \tau)}{q v \tau} dv = \left(\frac{Z+1}{Zq\bar{v}\tau}\right) \frac{\text{sin}(Z \text{tan} ^{-1}\Lambda)}{(1+\Lambda ^2)^{Z/2}},
\end{equation}
where $\Lambda=(q \bar{v} \tau)/(Z+1)$.

Figure \ref{fig:example_ISF} (green curve) shows an example of an ISF calculated at $q=1\;\mu{\rm m}^{-1}$ using typical {\em E. coli} motility parameters in Eqs.~\ref{eq:DDM2}-\ref{eq:analytical_integral}. The ISF shows a characteristic two-stage decay. The integral in Eq.~\ref{eq:DDM2} due to the straight-line motion of swimmers dominates the first, faster, process, while the purely diffusive first term due to the Brownian motion of non-swimmers dominates the second, slower, process.

Much can be learnt from visual inspection of this $f(q,\tau)$. The relative amplitudes of the fast and slow processes can easily be estimated to be $\approx 7:3$, which gives an estimated $\alpha \approx 0.7$. The length scale probed at this $q$ is $\ell \sim 2\pi/q \sim 6\;\mu$m. Either by extrapolating the green curve or by reference to the red curve for pure swimmers, it can be estimated that the fast process decays completely in $\tau_{\rm swim} \sim 0.5$~s. An order of magnitude estimate of the swimming speed is therefore $v \sim \ell/\tau_{\rm swim} \sim 12\;\mu{\rm m/s}$. The slower, diffusive, process decays completely in $\tau_{\rm diff} \sim 20$~s, so that an estimate of the diffusion coefficient of the non-swimmers can be obtained from $6D\tau_{\rm diff} \sim \ell^2$, giving  $D \sim 0.35\;\mu{\rm m}^2{\rm /s}$. These are credible estimates of the parameters used to generate this ISF: $v = 15\;\mu{\rm m /s}$, $D = 0.3\;\mu{\rm m}^2{\rm /s}$, and $\alpha = 0.7$.
\subsection{DDM Results}
\begin{figure}
   \begin{center}
      \includegraphics*[width=3.25in]{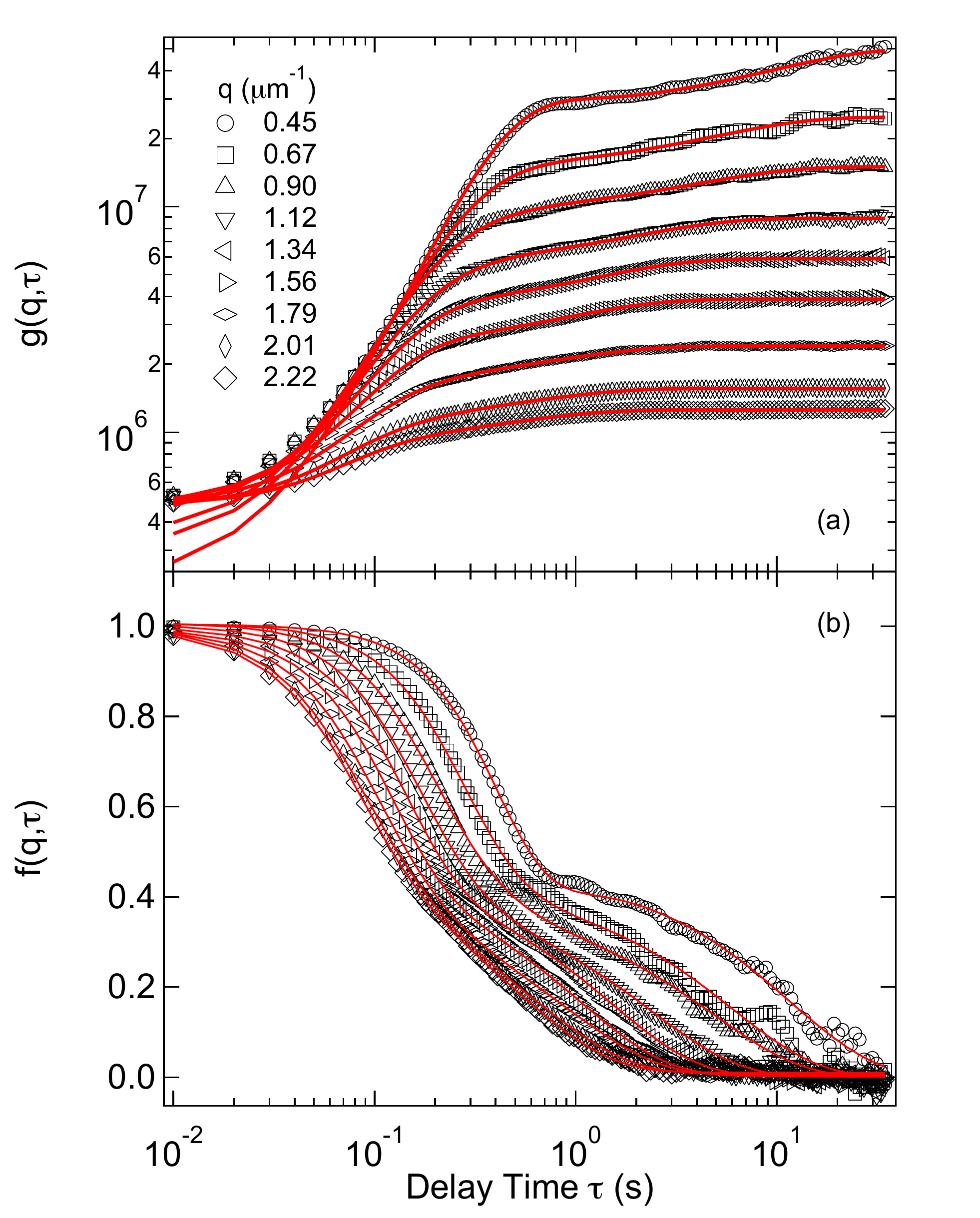}
      \caption{DDM for smooth swimming {\em E. coli}. (a) Measured (symbols) and fitted (lines) DICFs, $g(q,\tau)$. (b) ISFs, $f(q,\tau)$, reconstructed from $g(q,\tau)$ using Eqs.~\ref{eq:g_f},\ref{eq:DDM2} $\&$ \ref{eq:analytical_integral}.}
      \label{fig:DICF_ISF}
   \end{center}
\end{figure}

Fig.~\ref{fig:DICF_ISF}a shows typical DICFs, $g(q,\tau)$, measured using DDM in the range $0.45 \leq q \leq 2.22\;\mu {\rm m}^{-1}$ for a  suspension of SS {\em E.coli} mutant  $\Delta${\em cheY}. The measured $g(q,\tau)$ have a characteristic shape reminiscent of the calculated $f(q,\tau)$ shown in Fig.~\ref{fig:example_ISF} (Note the log-scale for the y-axis in Fig.~\ref{fig:DICF_ISF}a); indeed, Eq.~\ref{eq:g_f} shows that $g(q,\tau)$ should take the shape of an (un-normalized) `up-side-down' $f(q,\tau)$. Eq.~\ref{eq:g_f} also shows that the value of $g(q,\tau)$ at small $\tau$ gives a measure of the camera noise parameter $B(q)$, which is therefore seen to be more or less $q$-independent. The total amplitude of $g(q,\tau)$ measures the pre-factor $A(q)$, which evidently increases rapidly as $q$ decreases. This reflects the strong $q$ dependence of both the form factor of a single bacterium and the contrast function of the microscope objective.

The above qualitative remarks can be quantified by fitting the measured $g(q,\tau)$ using Eqs.~\ref{eq:g_f}, \ref{eq:DDM2} $\&$ \ref{eq:analytical_integral}. From the fit, we extract the six parameters $\bar v$, $\sigma$, $D$, $\alpha$, $A$ and $B$. 
The fitted functions $A(q)$ and $B(q)$ allow us to calculate $f(q,\tau)$ from the measured $g(q,\tau)$ via Eq.~\ref{eq:g_f} \cite{Note_1}, Fig.~\ref{fig:DICF_ISF}b.

\begin{figure}
   \begin{center}
      \includegraphics*[width=3.25in]{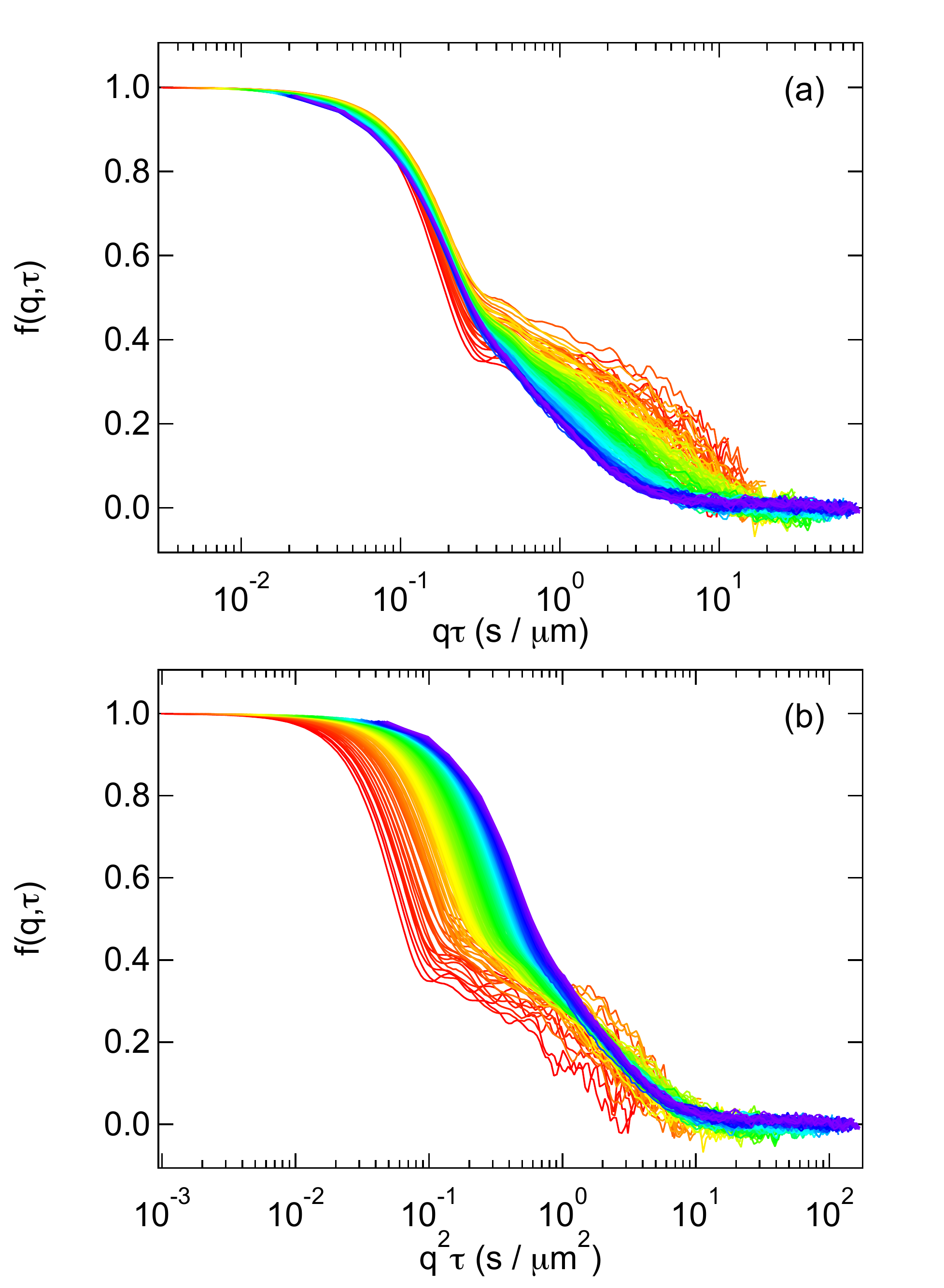}
      \caption{The reconstructed ISFs, $f(q,\tau)$, shown in Fig.~\ref{fig:DICF_ISF} plotted against (a) $q\tau$ and (b) $q^2\tau$. Data collapse for the fast process in (a) and for the slow process in (b). $q$ value increases from red to blue end of the spectrum colour in the range $0.3 \leq q \leq 2.2\;\mu \text{m}^{-1}$.}
      \label{fig:ISF_qt}
   \end{center}
\end{figure}

The ISFs calculated from experimental data (especially those for $q\approx 1\;\mu{\rm m}^{-1}$) show the characteristic shape already encountered in the theoretical ISF shown in Fig.~\ref{fig:example_ISF}: a fast decay due to swimming, followed by a slow decay due to diffusion. The identity of these two processes is confirmed by the different scaling of the time axis required to collapse the data at different $q$ values: the slow (diffusive) decay scales as $q^2\tau$, Fig.~\ref{fig:ISF_qt}a, and the fast (swimming, or ballistic) decay scales as $q\tau$, Fig.~\ref{fig:ISF_qt}b.

A clear separation of the swimming and diffusive decays is important for robust fitting of the ISF using Eq.~\ref{eq:DDM2}. Such separation of time scales will be achieved if a cell takes much less time to swim the characteristic distance probed, $\ell = 2\pi/q$, than to diffuse over the same distance (in the image plane), i.e. $\tau_{\rm swim} \sim \ell/v \ll \tau_{\rm diff} \sim \ell^2/4D$, which requires $q \ll q_c \sim v/D \sim 20-50\;\mu{\rm m}^{-1}$ for typical {\em E. coli} values of $v$ and $D$. All the data shown in Fig.~\ref{fig:DICF_ISF}b fit comfortably into this regime \cite{regime}.

\begin{figure}
   \begin{center}
      \includegraphics*[width=3.25in]{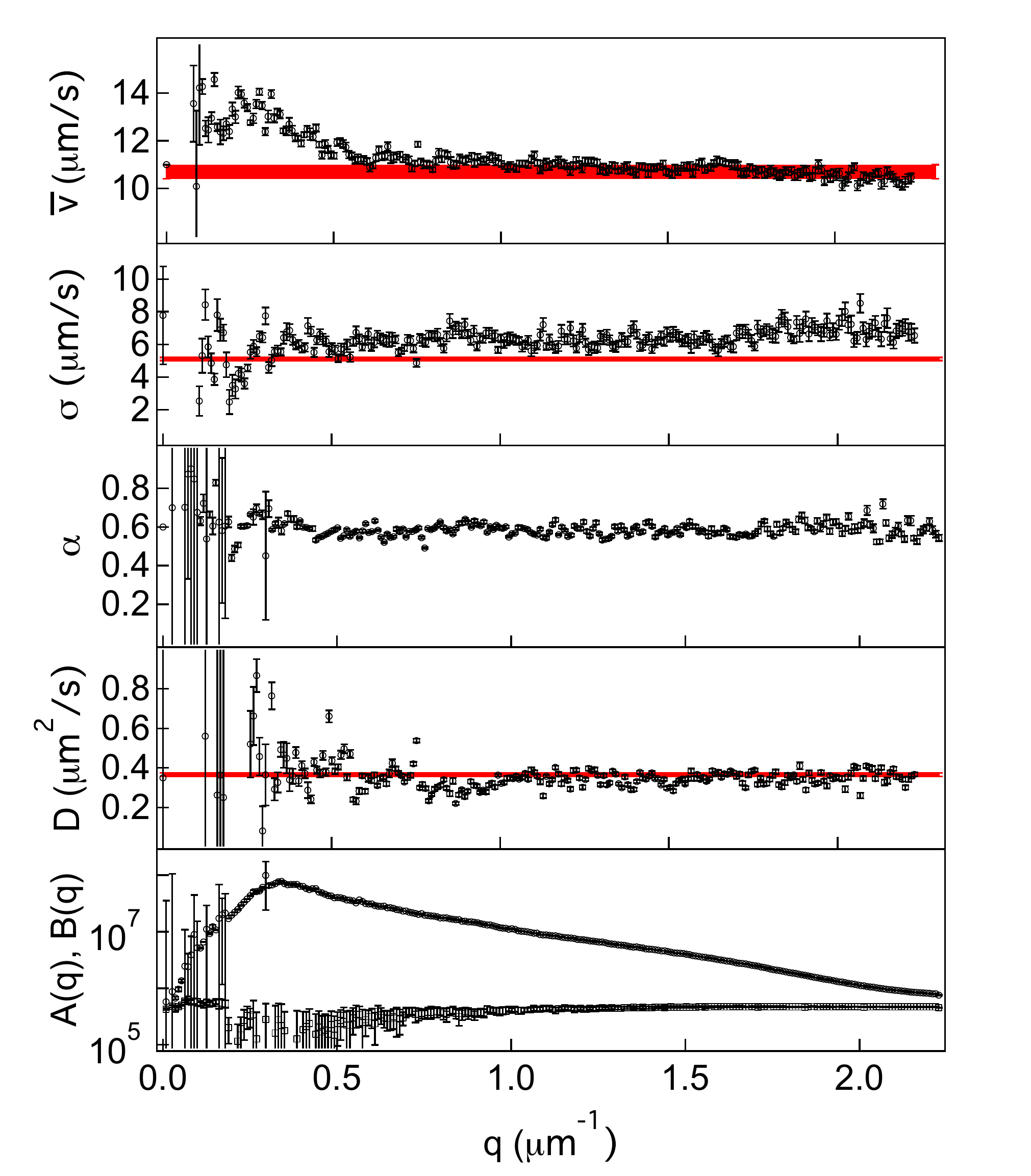}
      \caption{SS {\it E. coli}. Fitting parameters vs $q$ using Eqs.~\ref{eq:g_f}, \ref{eq:DDM2} $\&$ \ref{eq:analytical_integral}. From top to bottom: $\bar{v}$ and $\sigma$ of the Schulz distribution, motile fraction $\alpha$, diffusivity $D$, and $A(q)$ ($\circ$) and $B(q)$~($\square$). Red lines are results from tracking, with thickness corresponding to the error bars. No reliable value for $\alpha $ could be obtained from tracking.}
      \label{fig:Fit_parameters}
   \end{center}
\end{figure}

Fig.~\ref{fig:Fit_parameters} shows the fit parameters $(\bar{v}, \sigma, \alpha, D, A, B)$ from Eq.~\ref{eq:DDM2}-\ref{eq:analytical_integral} as functions of $q$. A common feature, particularly evident in $D(q)$, is the enhanced noise at low $q$. This is because at low $q$, the long-time, diffusive part of $f(q,\tau)$ has not fully decayed yet in our time window, rendering it harder to determine $D$ accurately. This can be improved by probing $g(q,\tau)$ over long times. To within experimental uncertainties, the motility parameters $(\bar{v}, \sigma, \alpha, D)$ are all $q$ independent for $q\gtrsim 1\;\mu m^{-1}$, which suggests that our model, Eq.~\ref{eq:DDM2}, is indeed able to capture essential aspects of the dynamics of a dilute mixture of non-motile and smooth swimming {\em E. coli}. Note that a fit using fixed $D$, over the full $q$-range, results in $q$-independent motility parameters only when the value used for $D$ is within $10\%$ of the value found in free fitting (data not shown). Averaging over $q$, in the range $0.5 \lessapprox q \lessapprox 2.2$, yields $\bar{v}=10.9 \pm 0.3\;\mu {\rm m/s}$, $\sigma=6.43 \pm 0.04\;\mu {\rm m/s}$, $\alpha=0.585 \pm 0.002$ and $D=0.348 \pm 0.003\;\mu {\rm m}^2{\rm /s}$ with error bars being the standard deviation of the mean in all cases except for $\bar{v}$ where estimated error bars reflects the residual $q$ dependence.

Note that the value of $D=0.35\;\mu {\rm m}^2{\rm /s}$ is higher than the value \cite{Wilson2011} of $D=0.30\;\mu {\rm m}^2{\rm /s}$  expected for a suspension of purely non-motile {\em E. coli} with "paralyzed" flagella ($\Delta${\em motA}) with similar geometry as the wild type. This is due to an enhancement of the diffusion of non-motile cells (and other colloidal-sized objects) in a suspension of motile organisms \cite{Wilson2011,Clement2011}. Moreover, the fitting of $D$ is dominated by the diffusion of non-motile organisms: changing our model from Eq.~\ref{eq:DDM2} to one in which the motile cells do not diffuse does not change the results (data not shown).

\begin{figure}
   \begin{center}
      \includegraphics*[height=3.25in]{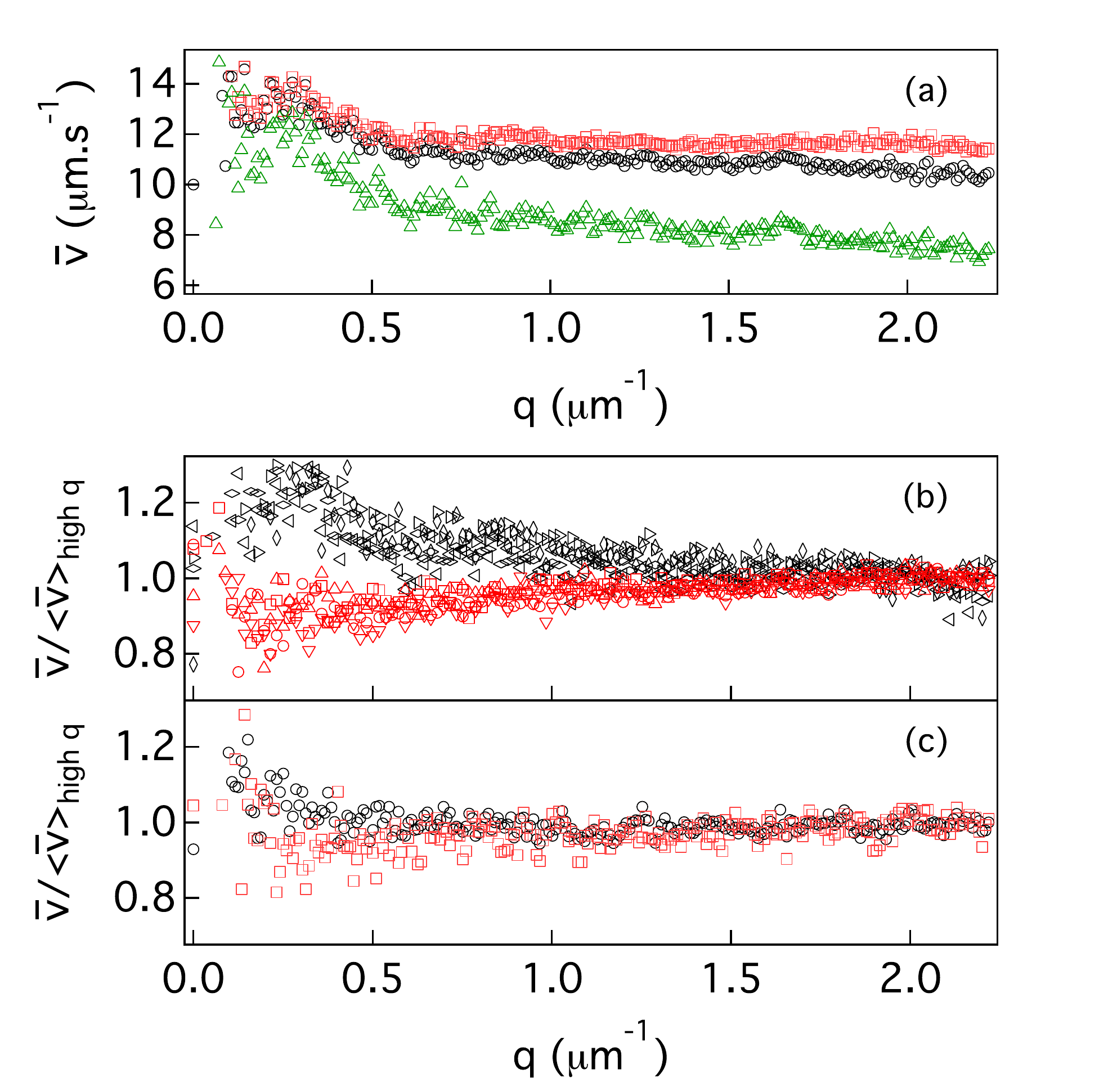}
      \caption{Swimming speed vs. $q$ from DDM. (a) Effect of using different forms of the speed distribution: lognormal ({\color{red} $\square$}), Schulz ($\circ$), and Gaussian ({\color{green}$\triangle$}). (b) Effect of tumbling (experiments): four data sets from the SS (black symbols) and four data sets from the WT (red symbols). (c) Effect of tumbling (simulations): SS ($\circ$) and WT ({\color{red}$\square$}). Note that for (b) and (c) panels, the swimming speed has been normalised to $\langle v \rangle_{\text{high q}}$ to enable comparison and highlight the difference in $q$-dependence.}
       \label{fig:Fusion}
   \end{center}
\end{figure}

We used a Schulz distribution for modelling $P(v)$ for analytic convenience in the integration of Eq.~\ref{eq:DDM2}. In Fig.~\ref{fig:Fusion}a we show the average speed obtained by fitting with three different probability distributions. The results for the Schulz and log-normal distributions agree closely, but using a Gaussian form produced noisier data and a significantly lower $\bar{v}$. The latter is because $P(v =0) \neq 0$ for the Gaussian distribution, strongly overestimating the number of slow swimmers. The presence of these spurious slow swimmers in turns renders it difficult to fit $D$, causing noisier data for all motility parameters.

We were able to fit the data satisfactorily irrespective of whether bright field, phase contrast or fluorescence imaging was used. However, phase-contrast imaging shows a better signal to noise ratio ($A(q)/B(q)$). In particular, changing $A(q)$ and $B(q)$ by using a $20 \times$ phase-contrast objective (which is suboptimal for our experiment) produced the same results in the relevant $q$ range (data not shown).
\subsection{Tracking results}
\label{subsec:tracking}
\begin{figure}
   \begin{center}
      \includegraphics[height=3.8in]{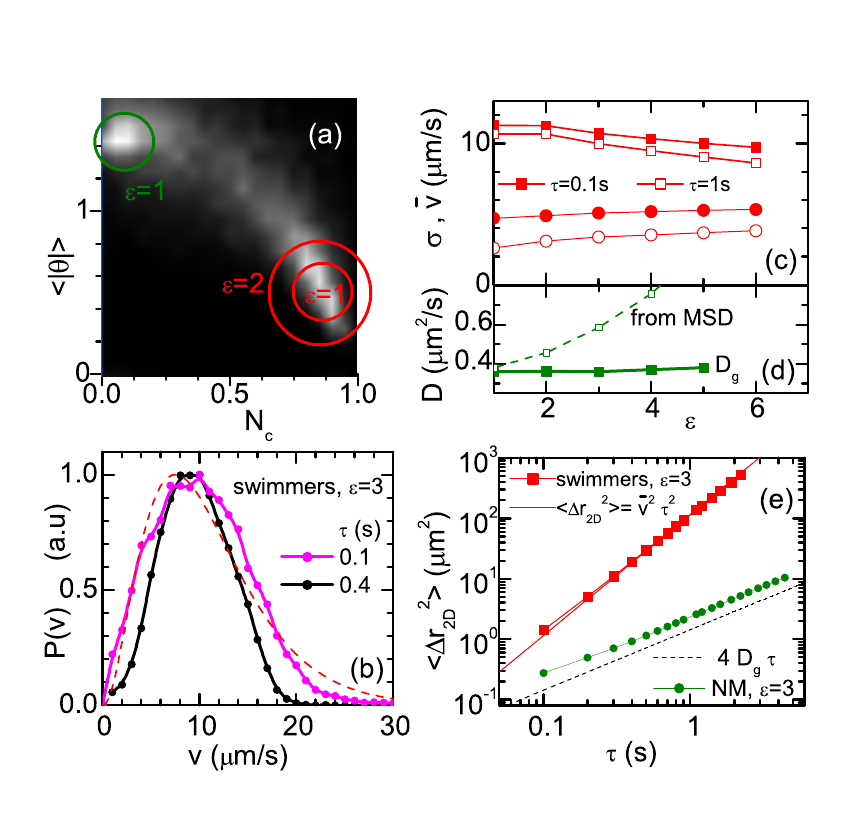}
      \caption{Tracking of SS {\it E. coli}. (a) Probability $P(N_c,\langle |\theta| \rangle)$ ($\Delta t=0.1~$s) for all tracks. White denotes large values of $P$. Circles (radius $\varepsilon$, see text) are selection criteria for motile (red, lower right) or NM (green, upper left) cells. (b) $P(v)$ for $\varepsilon=3$ ($\Delta t=0.1~$s) for two time lags $\tau$. Dashed line: Schulz distribution from DDM. (c) $\bar{v}$ (squares) and $\sigma$ (circles) of $P(v)$ vs $\varepsilon$ for $\tau=0.1$~s (filled) and $\tau=1$~s (open). (d) Diffusion coefficient of NM cells vs. $\varepsilon$, from Gaussian fits to $P(\Delta x_{NM}(\tau))$ ({\color{green} $\blacksquare$}) and from linear fits to the MSD ({\color{green} $\square$}). (e) MSD vs. $\tau$ for motile ({\color{red} $\blacksquare$}) and NM cells ({\color{green} $\bullet$}) for $\varepsilon=3$, $\Delta t = 0.1$~s. Line: motile MSD calculated using ${\bar v}_{\varepsilon=3}$ in (c); dashed line: NM MSD calculated using $D_{g,\varepsilon=3}$ from (d).}
      \label{fig:ecoli}
   \end{center}
\end{figure}

Figure~\ref{fig:ecoli}(a) show the probability density of the track diagnostics $(N_c,\langle |\theta| \rangle)$ (Section \ref{trackmethod}). Recall that $(N_c,\langle |\theta| \rangle)=(1,0) $ for straight swimming and $(0,\pi/2)$ for Brownian diffusion. While two clear maxima corresponding to diffusion and (nearly-straight) swimming are observed, there is a substantial statistical weight of tracks with intermediate $(N_c,\langle |\theta|\rangle)$ values. The actual distribution obtained depends on $\Delta t$, the elementary time interval into which we segment trajectories. Our optimal choice, $\Delta t = 0.1$s, over which the average swimming distance is $\approx 1~$pixel, gave the most sharply separated peaks. However, the motile and NM populations are still not cleanly separated in our $(N_c,\langle |\theta|\rangle)$ data, Fig.~\ref{fig:ecoli}a. We therefore select various populations of motile and NM cells by including tracks with $(N_c,\langle |\theta|\rangle)$ values within progressively larger circles centered on their respective peaks in the $(N_c,\langle |\theta|\rangle)$ space. The radius of the circle ($\epsilon$) is measured in units such that the $(0 \leq N_c \leq 1,0 \leq \langle |\theta|\rangle \leq \pi/2)$ space in Fig.~\ref{fig:ecoli}a is a $10 \times 10$ rectangle.

For motile cells, $P(v)$ was determined, at each $\epsilon$, by calculating the speed, $v=\langle \Delta r_{2D}(\tau)/\tau\rangle_T$, for each trajectory, averaged over the trajectory duration $T$, for various $\tau$. The limit $\tau \rightarrow 0$ gives the instantaneous linear velocity. In practice, the lowest reasonable $\tau$ is set by $\Delta t = 0.1$s. Figure~\ref{fig:ecoli}b shows $P(v)$ at $\epsilon=3$ for $\tau=0.1$~s and 0.4~s, while Fig.~\ref{fig:ecoli}c shows $\bar{v}$ and $\sigma$ of $P(v)$ for $\tau=0.1~$s and $\tau=1~$s. Unsurprisingly, $\bar{v}$ decreases with $\epsilon$ as progressively more `non-ideal' swimming tracks are included: first more curved trajectories and then (at larger $\epsilon$), some diffusive ones. Thus, there are ambiguities involved in motility characterisation using tracking. It was also not possible to extract reliably a value for $\alpha$ due to strong dependence on $\epsilon$.

However, the results for the other motility parameters show reasonable agreement with DDM (Fig.~\ref{fig:Fit_parameters}). In particular, using $(\Delta t,\tau) = (0.1{\rm s}, 0.1{\rm s})$ ({\color{magenta} $\bullet$}, Fig.~\ref{fig:ecoli}c), and averaging over all $\varepsilon$, $\bar{v} =10.7\pm 0.3\;\mu{\rm m/s}$ and $\sigma=5.1\pm0.1\;{\rm \mu m/s}$. The mean speed $\bar{v}_{\varepsilon}$ for each $\varepsilon$ is also consistent with the MSD of the swimmers, Fig.~\ref{fig:ecoli}e. The measured $P(v)$ depends on $\tau$, e.g. some fast swimmers will not be tracked for large $\tau$ unless perfectly aligned with the image plane, while for very short $\tau$ the 2D-projection contributes to $P(v)$ at small $v$ \cite{trackfoot}. Yet, for $\tau\sim 0.1-0.2$s, our measured $P(v)$ agrees with the Schulz distribution inferred from DDM, Fig.~\ref{fig:ecoli}c.

For NM cells, we determined $D$ by fitting the MSDs for selected tracks at several $\epsilon$. We again found a dependence on $(\Delta t,\tau)$ and $\epsilon$. The MSD for $\epsilon>1$ showed deviations from purely diffusive behaviour and/or the resulting values of $D$ depend significantly on $\varepsilon$ (Fig.~\ref{fig:ecoli}d, {\color{green} $\square$}). Both effects are due to (likely artificial) non-Gaussian tails in $P(\Delta x_{NM}(\tau))$ (not shown). Another estimate of the diffusion coefficient is $D_g$ (based on Gaussian fits to $P(\Delta x_{NM}(\tau))$, Section~\ref{trackmethod}), shown as a function of $\varepsilon$ in Fig.~\ref{fig:ecoli}d. The average value $D_g= 0.36\;\mu{\rm m}^2 {\rm /s}$ agrees with the DDM value of $0.35\;\mu{\rm m}^2 {\rm /s}$ and to the one from the MSD for $\varepsilon=1$. Figure.~\ref{fig:ecoli}e shows both the incorrect MSD of the diffusers obtained for $\varepsilon=3$ and the appropriate MSD based on $D_g$.
\subsection{Vertical motion and depth of field}\label{section:DF}
Our derivation of Eq.~\ref{eq:g_f} assumes that the image contrast of a bacterium does not vary with its position along the vertical (optical) $z$-axis, i.e. assumes an infinite depth of field ($\delta$). The validity of this assumption depends on how fast cells move relatively to the finite $\delta$ in reality. Giavazzi et al. presented a complex theoretical model, based on the coherence theory, to take into account this effect \cite{Giavazzi2009}. Here, we suggest a simple model and use simulations to investigate this effect and its importance over the accessible q-range. Our simple model captures the essential features and reproduces qualitatively and quantitatively the experimental results.

\begin{figure}
   \begin{center}
      \includegraphics*[width=3.25in]{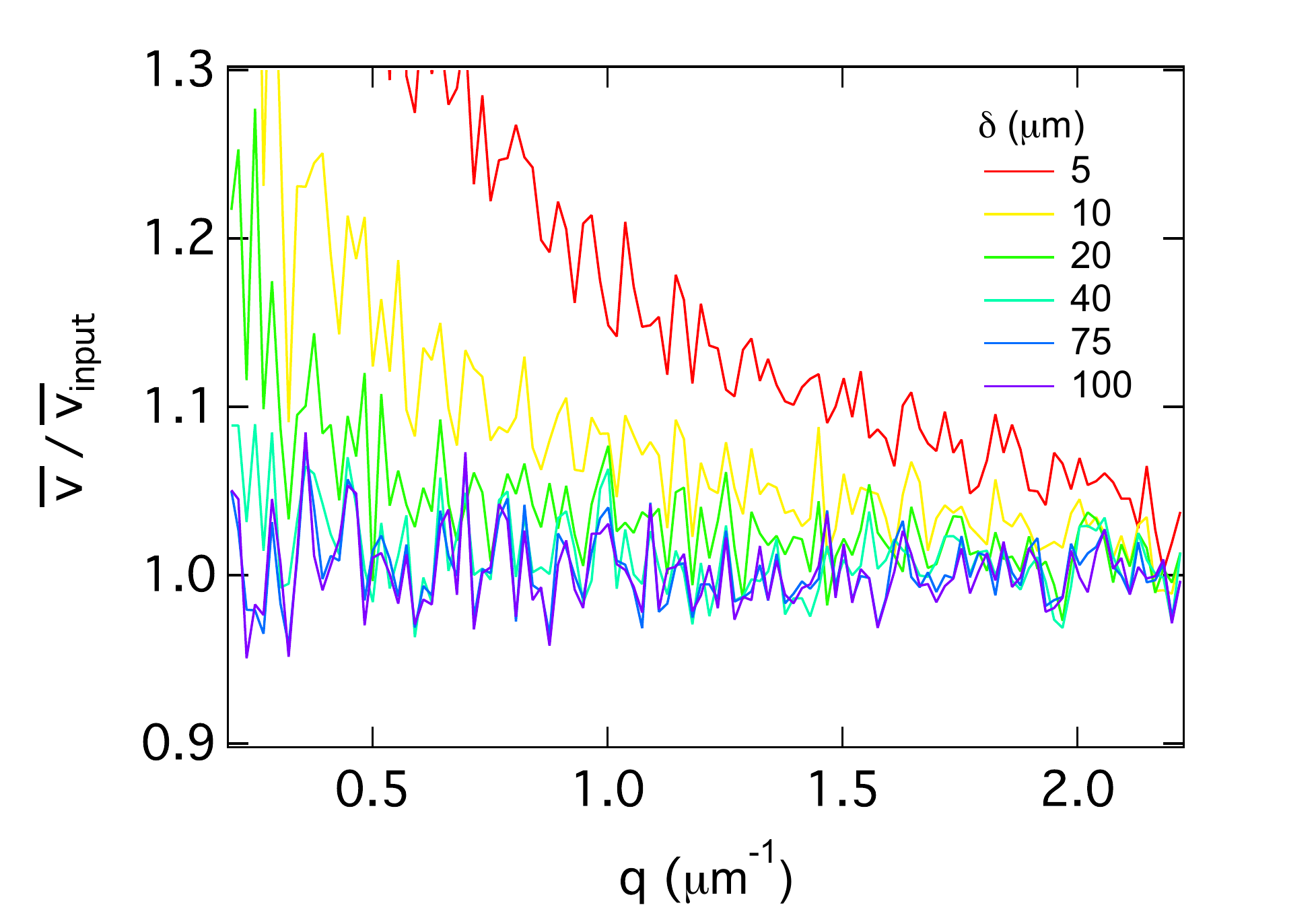}
      \caption{Effect of the depth of field on DDM analysis of simulated data. Normalised mean swimming speed $\bar{v}/\bar{v}_{input}$ versus $q$ for smooth swimmers and several values of depth of field $\delta$. $\bar{v}_{input}=15 \mu m/s$ is the input mean swimming speed used to generate the simulated data.}
      \label{fig:Fig8}
   \end{center}
\end{figure}

Experimentally \cite{Wilson2011}, the intensity profile of a bacterium along the $z$-axis can be described by the contrast function
\begin{equation}
C(z)= C_B- C_0 \left(1-\frac{4z^2}{\delta^2}\right)
\end{equation} \label{eq:contrast}
where $C_B$ and $C_0$ are the background and the amplitude of an object in the focal plane ($z=0$) respectively. We determined $C_B$ and $C_0$ experimentally, and then used this function to `smear' the simulated data previously presented \cite{Wilson2011} to give simulated `images' at a range of $\delta$. At each $\delta$ except the lowest, the input values $\{\bar{v}, \sigma, \alpha, D\}$ are recovered from DDM analysis of these images at $q \gtrsim 2\pi/ \delta$; the case of $\bar{v}$ is shown in Fig.~\ref{fig:Fig8}. However, for $q \lesssim 2\pi/\delta$, the analysis returns $\bar{v}$ and $D$ values that are too high: disappearance of cells along the $z$ axis due to the rapid fading of $C(z)$ is mistaken as swimming and diffusion. Comparison between simulated data (Fig.~\ref{fig:Fig8}) and experimental data (Fig.~\ref{fig:Fit_parameters} $\&$ \ref{fig:Fusion}b) shows that the effect of finite depth of field, $\delta$, is negligible for $q>0.5\;\mu \text{m}^{-1}$ using 10x phase-contrast imaging and that our experimental depth of field $\delta \gtrsim 20\;\mu$m.
\section{Wild-Type E.coli} \label{sec:wildtype}
The motility pattern of WT {\em E. coli} in the absence of chemical gradients is well known \cite{Berg1972}. A cell alternates between running for $t_{\rm run}\approx$~s and tumbling for $t_{\rm tum}\approx$~s. During the latter they change direction abruptly. After many tumbling events, the bacterium effectively performs a 3D random-walk.

Modelling the ISF using Eq.~\ref{eq:DDM2} assumes that swimmers follow straight trajectories and neglects the effect of tumbling. We have previously applied our method to the WT $E.\;coli$ \cite{Wilson2011}. Here we study the effect of tumbling by comparing systematically the $q$-dependence of the average speed obtained from DDM for WT (run and tumble) and SS (run only) swimmers. Since several experimental data sets were obtained from different batches of cells, we report the speed normalised to $\langle \bar{v} \rangle_{\text{high q}}$, the average in the range ($2.0<q<2.2\;\mu \text{m}^{-1}$).

Simulations and experiments show, Fig.~\ref{fig:Fusion}, a qualitative difference in $\bar{v}(q)$ of SS and WT cells. All data for WT cells show a slight {\em decrease} in $\bar{v}(q)$ towards low $q$, while data for SS show the opposite trend. The increase towards low $q$ in the $\bar{v}(q)$ of the SS is presumably largely due to depth of field effects (Section~\ref{section:DF}). The opposite trend in the behaviour of $\bar{v}(q)$ for the WT can be understood as follows. The mean speed, $\bar{v}(q)$, measured by DDM at a certain $q$ is estimated by $\bar{v}(q) \sim (2\pi/q)/\tau_q$, i.e. the time ($\tau_q$) taken to advect density between two points spatially separated by distance $2\pi/q$. For a straight swimmer, the tracklength $s$ will be equal to the distance between the two points, i.e. $s = 2\pi/q$, so that $\bar{v}(q)=\bar{v}$. Any deviation from a straight track, e.g. due to changes in direction from tumbling, renders $s > 2\pi/q$. Since $\tau_q = s/\bar{v}$, we now have $\bar{v}(q) < \bar{v}$. This effect  becomes progressively more pronounced at low $q$, as observed.

\section{Swimming algae: wild-type C. \protect \emph{reinhardtii}} \label{sec:algae}
As a final example we apply DDM to the bi-flagellate freshwater alga  {\it Chlamydomonas reinhardtii}, a model for eukaryotic flagellar motility \cite{Harris09}. {\it C. reinhardtii} has a prolate spheroidal cell body about $10\;\mu$m across with two flagella roughly $10$-$12 \;\mu$m long \cite{Harris09}. Beating the latter at $\simeq 50$Hz in an alternation of effective (forward moving) and recovery (backward moving) strokes propels the cell body forward, oscillating as it advances. The flagellar beat is not perfectly planar, so cells precess around their long axis at $\simeq 2$Hz; this rotation, critical to phototaxis \cite{FosterSmyth80}, results in helical swimming trajectories. For length scales $\sim 100\;\mu m$ the direction of the axis of the helical tracks is approximately straight, but on larger scales the stochastic nature of the flagellar beat causes directional changes resulting in random walk~\cite{Drescher2009,HillPedley05, Polinetal09}.

Racey {\it et al.} \citep{RaceyHallettNickel81} carried out the first high-speed microscopic tracking study of {\it C. reinhardtii}, obtaining the cells' swimming speed distribution, with mean speed $\bar v=84\,\mu$m/s, as well as the average amplitude $\bar A=1.53\,\mu$m and frequency $\bar f=49$ Hz of the beat. A more recent tracking study obtained a 2D swimming speed distribution with $\bar v\approx100\,\mu$m/s \citep{LeptosGollubPRL09}. The swimming of {\it C. reinhardtii} has also been studied with dynamic light scattering (DLS) \citep{RaceyHallettNickel81, RaceyHallett83}. However, studies at appropriately small $q$ are difficult, severely limiting useful data (the limitation is even greater than for bacteria: algae swim on larger length scales, requiring smaller values of $q$). We present here the first characterisation of the swimming motility of {\it C. reinhardtii} using DDM. As for {\it E. coli}, the technique allows to study the 3D motion of much larger numbers of cells ($\sim 10^4$) than tracking, and allows easy access to larger length scale than DLS.
\subsection{Model of $f(q,\tau)$}
The swimming dynamics of {\it C. reinhardtii} are on larger length scales, shorter time scale (algae swim faster) and of a different nature than {\it E. coli}, so models and experimental conditions used for DDM need to be adjusted for this organism. In particular, the decay of the ISF, $f(q,\tau)$, will reflect these algae's peculiar dynamics. Cells oscillate at lengthcales $<10 \,\mu m$, translate in the range  $10 \,\mu m<L<30 \,\mu m$, spiral over $30 \,\mu m<L<100 \,\mu m$, and diffuse for $L>100 \,\mu m$. A schematic representation of a helical trajectory, highlighting the small scale oscillatory motion is shown in inset of Fig.~\ref{fig:chlamytraj} (diffusive length scales are not shown).

At length scales $L\lessapprox 30\;\mu \text{m}$, the swimming of {\it C. reinhardtii} can be approximated as a sinusoidal oscillation superimposed on a linear progression, so that the particle displacement $\Delta r(\tau)$ of a cell after a time interval $\tau$ is given by \citep{RaceyHallett83}
\begin{equation}\label{eq:algae}
\Delta r(\tau)=v \tau + A_0[\sin(2\pi f_0 \tau+\phi)-\sin(\phi)],
\end{equation}
where $A_0$ and $f_0$ are the amplitude and frequency of the swimming oscillation and $\phi$ is a random phase to ensure the swimming beats of different cells are not synchronised. Substituting this into Eq.~\ref{fqt}, averaging over $\phi$, and assuming a Schulz distribution for the swimming speed (Eq.~\ref{eq:schultz}), we obtain the ISF,
\begin{equation}\label{eq:DDMalgae}
f_{\rm{algae}}(q,\tau)= \frac{1}{2}\int_{-1}^1 \frac{\cos\left[(Z+1) \tan^{-1}\left(\Lambda\chi\right)\right]}
{\left[1+\left(\Lambda\chi\right)^2\right]^{(Z+1)/2}} J_0[2 q A_0 \chi \sin(\pi f_0 \tau)]\,d\chi,
\end{equation}
where $\Lambda\equiv\frac{q \overline{v} \tau}{Z+1}$, $\chi\equiv \cos \psi$, where $\psi$ is the angle between ${\vec q}$ and ${\vec r}$, $J_0$ is the zeroth order Bessel function. All other terms are as previously defined. The first and second term describe the contribution from straight swimming and oscillatory beat, respectively. In the limit $q A_0\ll1$ ($J_0\to1$), Eq.~\ref{eq:DDMalgae} analytically integrates to Eq.~\ref{eq:analytical_integral}, the same expression as for the progressive model used for {\it E. coli}.

The derivation of Eq.~\ref{eq:DDMalgae} assumes the distributions $P(A)$ and $P(f)$ for swimming amplitude and frequency, respectively, are narrowly centred around the values $A_0$ and $f_0$ and ignores (i) the negligible diffusion of non-motile algae; (ii) any bias in the swimming direction caused by gravitaxis \cite{HillHader97}; and (iii) the helical nature of the swimming.
\subsection{DDM results}
\begin{figure}
   \begin{center}
      \includegraphics*[width=3.25in]{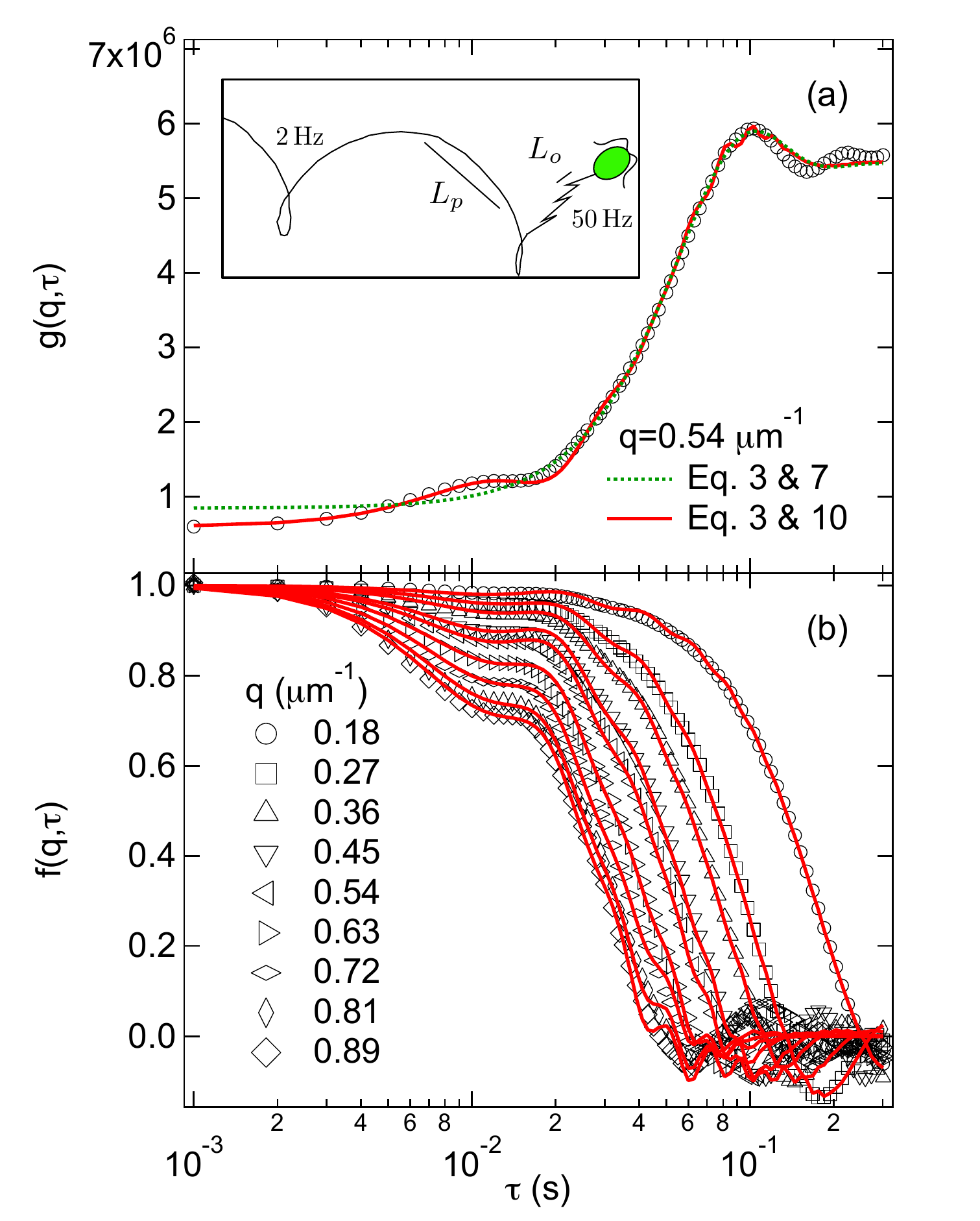}      
      \caption{DDM for WT C. {\it reinhardtii}. (a) Measured (circles) $g(q,\tau)$. Line and dashed line are fits using the oscillatory model (Eq. \ref{eq:DDMalgae}) and the linear model (Eq.~\ref{eq:analytical_integral}), respectively. Inset: Portion of a helical {\it C. reinhardtii} trajectory. The progressive, $L_p$, and (zoomed-in) oscillatory, $L_0$, length scales probed by DDM are shown, with the frequencies of the helical precession (2~Hz) and oscillatory swimming (50~Hz). (b) ISFs, $f(q,\tau)$, using Eqs.~\ref{eq:g_f} $\&$ \ref{eq:DDMalgae}.}
      \label{fig:chlamytraj}
   \end{center}
\end{figure}

Fig.~\ref{fig:chlamytraj}a shows a typical DICF, $g(q,\tau)$, at $q=0.52\;\mu {\rm m}^{-1}$ ($l\approx 12\;\mu \text{m}$), for a suspension of WT alga {\it C. reinhardtii} measured using DDM. The reconstructed ISFs, $f(q,\tau)$, are shown in Fig.~\ref{fig:chlamytraj}b in the range $0.2 \lessapprox q \lessapprox 0.9\;\mu \text{m}^{-1}$, corresponding to a length scale range of $7 \lessapprox l \lessapprox 30\;\mu \text{m}^{-1}$. $f(q,\tau)$ shows a characteristic shape for all $q$'s: a fast decay at $\tau \leq 0.02\;s$ due to the oscillatory beat and a slower decay at $\tau \geq 0.02\;s$ due to swimming. The identity of these two processes is confirmed by their difference in $\tau -$ and $q-$ dependencies. The characteristic time of the fast process is independent of $q$, while its amplitude decreases with $q$. Both observations fully agree with the term ($J_0$) due to the oscillatory contribution in Eq.~\ref{eq:DDMalgae}. Moreover, $0.02\;s$ corresponds to the period of a $50$~Hz oscillatory beat. Finally, the slow process scales perfectly with $q\tau$ (data not shown) confirming the ballistic nature (swimming) of this process.

\begin{figure}
   \begin{center}
      \includegraphics*[width=3.in]{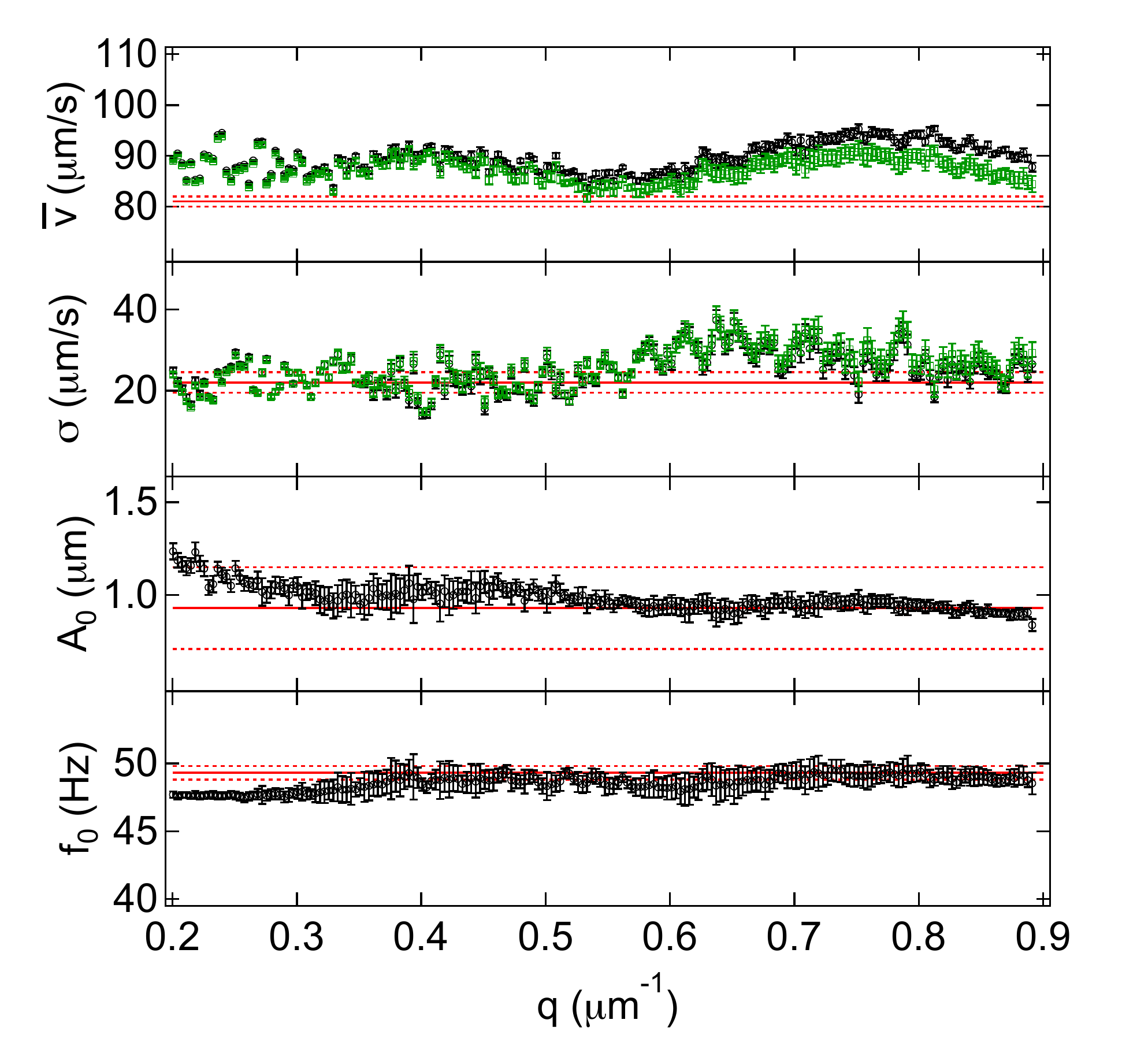}          
      \caption{Fitting parameters using the oscillatory model Eq.~\ref{eq:DDMalgae} (circles) or linear model Eq.~\ref{eq:analytical_integral} (squares) as function of $q$ for {\it C. reinhardtii}. From top to bottom: $\bar{v}$ and $\sigma$ of the Schulz distribution, amplitude $A_0$, and frequency $f_0$. Red lines are results from tracking with dashed lines corresponding to error bars.}
      \label{fig:vAfVSq-algae}
   \end{center}
\end{figure}

Fig.~\ref{fig:vAfVSq-algae} shows the fitting parameters $(\bar{v}, \sigma, A_0, f_0)$ from Eq.~\ref{eq:g_f} using the oscillatory model (Eq.~\ref{eq:DDMalgae}) as a function of $q$. All parameters display a small $q$-dependence. This is likely due to effects not captured by the simple oscillatory model (e.g. body precession and helical swimming) and will be discussed elsewhere. Averaging over $q$ yields $\bar{v}=89.6\pm 2.8\;\mu$m/s, $\sigma=24.9\pm4.6\;\mu$m/s, $A_0=0.98\pm0.06\;\mu$m and $f_0=48.6\pm0.55$ Hz, with estimated error bars reflecting the residual $q$ dependence. Fitting the experimental data using Eq.~\ref{eq:DDMalgae} requires numerical integration. Using instead the linear model (Eq.~\ref{eq:analytical_integral}), thus ignoring the oscillatory beat, yields similar results for ($\bar{v}$,$\sigma$) as shown in Fig.~\ref{fig:vAfVSq-algae}. This is because the fast process is mainly ignored when performing such fit as shown in Fig.~\ref{fig:chlamytraj}a. Moreover, using the linear model and a movie, for which the lowest $\tau \gtrapprox 1/f_0$ (for example 100 fps) and therefore the oscillatory beat is not contributing to the ISF, yields the same $(\bar{v},\sigma)$, thus allowing high-throughput economical measurements of the mean speed of biflagellate algae.
\subsection{Tracking results}
\label{sec:algaetrack}
\begin{figure}
   \begin{center}
     \includegraphics[width=3.25in]{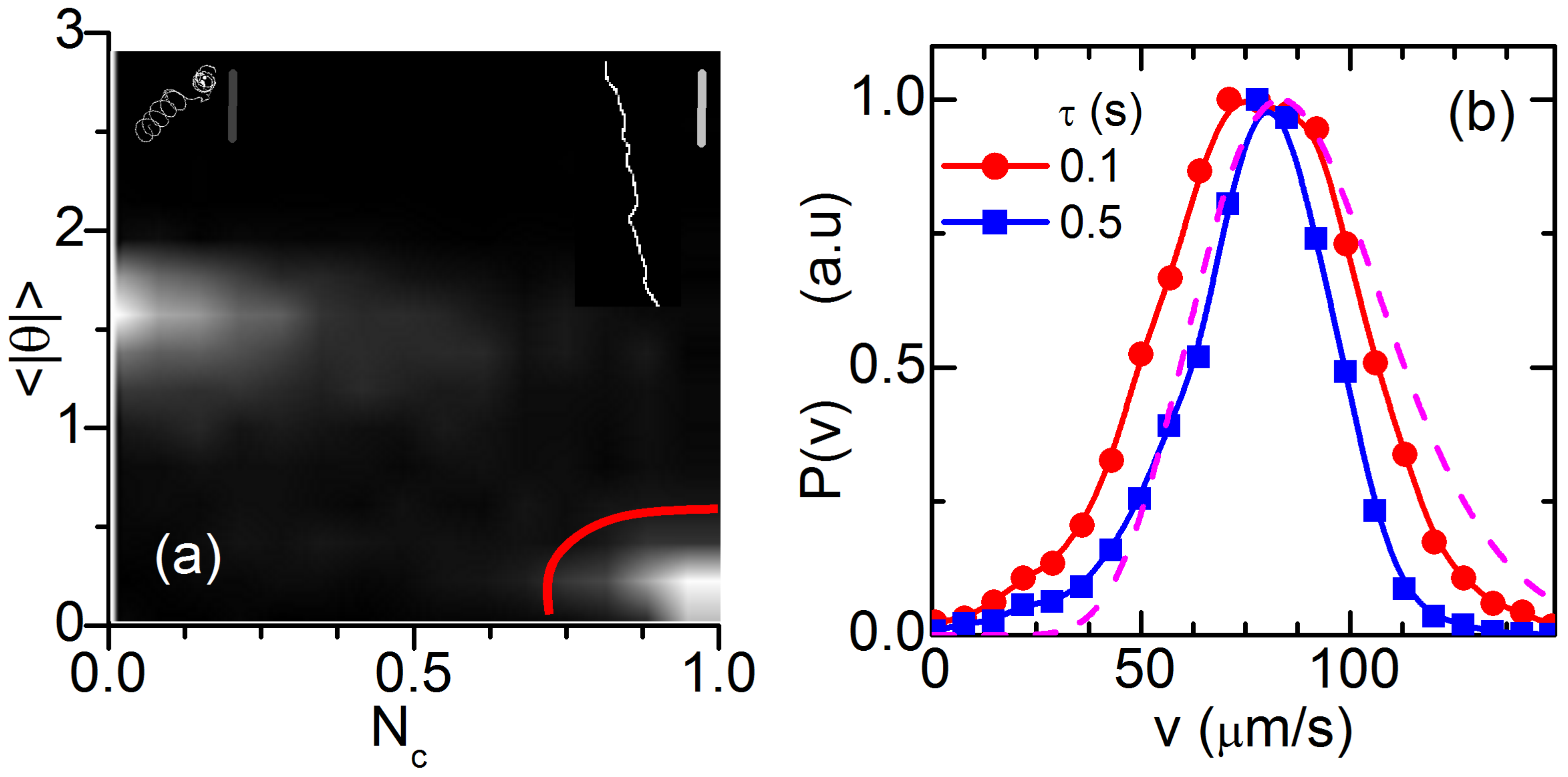}
      \caption{Tracking of {\it C. reinhardtii}. (a) Probability $P(N_c,\langle |\theta| \rangle)$ of all tracks ($\Delta t=0.05~$s). Tracks within the red bordered region (example in top right inset, $7~$s, scalebar $150~\mu$m) are used to measure $P(v)$; Top left inset: an excluded track with $N_c<0.4$ ($30~$s, scalebar $30~\mu$m). (b) Normalised $P(v)$ from tracks selected in (a), for two values of $\tau$. Dashed line: $P(v)$ from DDM analysis.}
      \label{fig:algae_ncth_pv}
   \end{center}
\end{figure}

Tracking of  {\it C. reinhardtii} resulted in two well separated groups of  $(N_c, \langle |\theta| \rangle)$ values (Section \ref{trackmethod}) independent of $\Delta t$, provided $\Delta t > 1/f_0$, Fig.~\ref{fig:algae_ncth_pv}a. We used tracks with $(N_c \ge 0.7$, $\langle |\theta| \rangle<0.5$), reflecting nearly straight swimmers aligned with the image plane, to obtain $P(v)$. Misaligned tracks are excluded this way: motion perpendicular to the helical axis enhances the circular contribution in the 2D projection, thus reducing $N_c$ and increasing $\langle |\theta| \rangle$ (insets to Fig.~\ref{fig:algae_ncth_pv}a).

We measured $P(v)$ for several $\tau$, Fig~\ref{fig:algae_ncth_pv}b, and found a slight $\tau$-~dependency, e.g. due to undetected fast swimmers for large $\tau$. Note that our $P(v)$ is smaller at small $v$ than in \cite{LeptosGollubPRL09}, where all {\em projected} trajectories were considered. Our small $v$ data are likely closer to the true distribution, due to our exclusion of misaligned tracks. Moreover, our $P(v)$ are in reasonable agreement with the result inferred from DDM (Fig~\ref{fig:algae_ncth_pv}b). We find $\bar{v}=81 \pm 1~\mu$m/s and $\sigma=22 \pm 3 ~\mu$m/s (averaged over different $\tau$), in excellent agreement with the values from DDM  (Fig.~\ref{fig:vAfVSq-algae}). Extending the selected trajectories to ($N_c \ge $0.55$,\langle |\theta| \rangle<0.7)$, changes $P(v)$ and $\bar{v}$ by less than $5\%$.
We analysed the oscillating component of the displacement, $r_{os}(t)$, for `straight' tracks \citep{footnote:beat}. From Fourier analysis of $r_{os}(t)$, we obtained $f_0=49.3\pm0.5$~Hz. We identified an additional modulation frequency of $\sim 10$ Hz (i.e. an extra peak at $f -f_0 \simeq 10~$Hz in the power spectrum of $r_{os}(t)$, to be discussed elsewhere). From the rms value of $r_{os}(t)$, we determined the average oscillation amplitude $A_0=\sqrt{2 \langle x^2_{os}+y^2_{os} \rangle}=0.93 \pm 0.22~\mu$m. These values are in agreement with previous \cite{RaceyHallettNickel81} and DDM results.

Thus, our results simultaneously validate the DDM methodology and the simple model (Eq. \ref{eq:algae} $\&$ \ref{eq:DDMalgae}) for the swimming of {\em C. reinhardtii}. Our method can therefore be used for the fast and accurate characterisation of the motility of large ensembles of this organism, and, potentially, of other algae.
\section*{Conclusions}
We have shown that DDM is a powerful, high-throughput technique to characterise the 3D swimming dynamics of microorganisms over a range of time scales and length scales ($\sim 3$ and $\sim 1$ order of magnitude respectively) simultaneously in a few minutes, based on standard imaging microscopy. The time scales and length scales of interest depend on the swimming dynamics of the microorganism, and are easily tuned by changing the frame rate or optical magnification respectively.

We have studied in considerable detail how to use DDM for characterising the motility of smooth-swimming (run only) and wild-type (run and tumble) {\it E.~coli}, as well as the wild-type alga {\it C. reinhardtii}. We validated the methodology using tracking and simulations. The latter was also used to investigate the effect of a finite depth of field and tumbling. Using DDM, we were able to extract (i) for {\it E. coli}: the swimming speed distribution, the fraction of motile cells and the diffusivity; and (ii) for {\it C. reinhardtii}: the swimming speed distribution, the amplitude and frequency of the oscillatory dynamics. In both cases, these parameters were obtained by averaging over many thousands of cells in a few minutes without the need for specialised equipment.

Further developments are possible. For {\em E. coli}, analytic expressions for $\bar{v}(q)$ taking into account either trajectory curvature due to rotational Brownian motion (smooth swimmers) or directional changes due to tumbling (wild type), can be derived. Fitting these expressions to data should then yield quantitative information on the respective features. For {\em C. reinhardtii}, the helical motion, the asymmetric nature of the swimming stroke, and the higher harmonics in the body oscillations observed by tracking could be explored theoretically and using DDM experiments. This will allow us to test simulations that use the method of regularised stokeslets to reproduce the fine details of the swimming of bi-flagellate algae \citep{OmalleyBees11}.

DDM is based on the measurement of the spatio-temporal fluctuations in intensity, and therefore does not require good optical resolution of the motile objects. Thus, DDM can probe a large field of view, yielding good statistics even under relatively poor imaging conditions. Moreover, DDM could also be used to probe anisotropic or asymmetric dynamics \citep{Reufer2012} of microorganisms. Finally, the analysis is not restricted to dilute suspensions and can be used to investigate the swimming dynamics at different time scales and length scales of the collective behaviour of populations. However, quantitative interpretation of the resulting data will require new models of the ISF.

With the availability of DDM, quantitative characterisation of motility can become a routine laboratory method, provided suitable theoretical models are available for fitting of the ISF. Even without such models, however, qualitative features of the measured ISF can still allow conclusions to be drawn and trends to be studied (e.g. the speeding up of the decay of the ISF almost invariably correspond to faster motion). DDM should therefore be a powerful tool in future biophysical studies of microorganismic locomotion.
\section*{Acknowledgment}
Authors acknowledge support from FP7-PEOPLE (PIIF-GA-2010-276190); EPSRC (EP/D073398/1); the Carnegie Trust for the Universities of Scotland; the Swiss National Science Foundation (PBFRP2-127867 and 200021-127192) and EPSRC (EP/E030173/1 and EP/D071070/1).

\begin{thebibliography}{61}
\bibitem{software}
All relevant softwares are available on request.
\bibitem{Note_1} The determination of $A(q)$ and $B(q)$ does not necessarily requires fitting of $g(q,\tau)$ and therefore a model for $f(q,\tau)$. They could simply be determined from the short and long time limit of $g(q,\tau)$ such as: $B(q)=f(q,\tau \rightarrow0)$ and $A(q)=f(q,\tau\rightarrow \infty)-B(q)$. However, the latter requires well-defined "plateau" at both short and long time regimes, which are not always observed depending on the value of $q$, and the experimental time window restricted: at short-time by the frame rate and at long-time by the duration of the movie.
\bibitem{footnote:beat}
We define $r_{os}(t)=(x_{os},y_{os})(t)=r_{2D}(t)-\langle r_{2D}(t) \rangle_{dt}$ with $dt=2/f_0$. $r_{os}(t)$ is well resolved due to  subpixel accuracy $\sim 0.2~\mu$m of the coordinates.
\end{thebibliography}
\end{document}